\documentclass[journal = jctcce, manuscript=article]{achemso}
\usepackage{dcolumn}
\usepackage{bm}
\usepackage{amsmath}
\usepackage{rotating}
\usepackage{longtable}
\usepackage{booktabs}
\usepackage{graphicx}
\usepackage{epstopdf}
\setkeys{acs}{articletitle = true}
\author{Andrew M. Sand}
\affiliation{Department of Chemistry, Chemical Theory Center, and the Minnesota Supercomputing Institute,
The University of Minnesota, Minneapolis, MN 55455 USA}
\author{Chad E. Hoyer}
\affiliation{Department of Chemistry, Chemical Theory Center, and the Minnesota Supercomputing Institute,
The University of Minnesota, Minneapolis, MN 55455 USA}
\author{Kamal Sharkas}
\affiliation{Department of Chemistry, Chemical Theory Center, and the Minnesota Supercomputing Institute,
The University of Minnesota, Minneapolis, MN 55455 USA}
\author{Katherine M. Kidder}
\affiliation{Department of Chemistry, Chemical Theory Center, and the Minnesota Supercomputing Institute,
The University of Minnesota, Minneapolis, MN 55455 USA}
\author{Roland Lindh}
\affiliation{Department of Chemistry - {\AA}ngstr\"{o}m, The Theoretical Chemistry Programme, Uppsala University, P.O. Box 518, SE-751 20
Uppsala, Sweden}
\author{Donald G. Truhlar}
\affiliation{Department of Chemistry, Chemical Theory Center, and the Minnesota Supercomputing Institute,
The University of Minnesota, Minneapolis, MN 55455 USA}
\email{truhlar@umn.edu}
\author{Laura Gagliardi}
\affiliation{Department of Chemistry, Chemical Theory Center, and the Minnesota Supercomputing Institute,
The University of Minnesota, Minneapolis, MN 55455 USA}
\email{gagliard@umn.edu}

\affiliation{Department of Chemistry, Chemical Theory Center, and the Minnesota Supercomputing Institute,
The University of Minnesota, Minneapolis, MN 55455 USA}

\title{Analytic Gradients for Complete Active Space Pair-Density Functional Theory}
\date{\today}

\begin{document}
\maketitle
\begin{abstract}
Analytic gradient routines are a desirable feature for quantum mechanical methods, allowing for efficient determination of equilibrium and transition state structures and several other molecular properties.  In this work, we present analytical gradients for multiconfiguration pair-density functional theory (MC-PDFT) when used with a state-specific complete active space self-consistent field reference wave function.  Our approach constructs a Lagrangian that is variational in all wave function parameters.  We find that MC-PDFT locates equilibrium geometries for several small- to medium-sized organic molecules that are similar to those located by complete active space second-order perturbation theory but that are obtained with decreased computational cost.
\end{abstract}

\section{Introduction}

Energy gradients play an important role in quantum chemistry.  They are needed to determine equilibrium structures, transition state structures, trajectories in reaction dynamics, and many molecular properties such as nuclear magnetic resonance shifts~\cite{Pulay,helgaker_1988}.  Gradients can be computed numerically~(by using finite differences) or analytically.  The numerical approach, however, requires a costly number of single-point calculations, and its accuracy must be checked with respect to the finite difference step size.  Thus, analytic approaches~\cite{Kato,rice_1986,helgaker_1989,shepard_1992,Grad_book,Stalring,jagau_2010,shiozaki_11,liu_2013,schutski_2014,bozkaya_2016,park_2017},
 although more mathematically-involved, are preferred due to lower cost and higher accuracy.
An accurate modeling of chemical systems requires quantum mechanical methods capable of describing static and dynamic correlation~\cite{Roos_book,Hirao_book,Gordon_book}.  Methods which employ a single Slater determinant often do not give satisfactory results for strongly-correlated systems; the use of multiconfiguration wave function approaches~\cite{szalay_2012} is desirable for such systems.  However, these methods usually scale poorly with system size.

Multiconfiguration pair-density functional theory\cite{mcpdft,MCPDFT_ACR}~(MC-PDFT) is an affordable and accurate type of density functional theory that adds dynamic correlation energy to multiconfiguration self-consistent field~(MCSCF) wave functions using a density functional called an on-top functional.  The total electronic energy is expressed as a sum of the MCSCF kinetic energy and a functional of both the one-electron density and the two-electron on-top pair density of a reference wave function.  Whereas the one-electron density is the probability of finding an electron at a point in space, the two-electron on-top pair density is related to the probability of finding two electrons on top of each other at a point in space~\cite{Carlson17,becke_1995,moscardo_1991,perdew_1995,gusarov_2004}.   

The advantage MC-PDFT has over other post-MCSCF methods is that MC-PDFT can calculate the dynamic correlation energy correction at a cost that scales as a single iteration of the reference MCSCF calculation~\cite{Sand2017}.  In contrast, post-MCSCF wave function methods such as complete active space second-order perturbation theory~(CASPT2)~\cite{andersson_1990}, multireference M{\o}ller-Plesset perturbation theory~(MRMP)\cite{hirao1,hirao2,hirao3} and multireference configuration interaction~(MRCI)~\cite{lischka_1981}, while often providing a good description of correlation energy for smaller-sized systems, scale poorly with system size, which limits applications to many medium and large-sized systems.  MC-PDFT has been shown to be able to predict energies with an accuracy similar to CASPT2 at a much lower computational cost~\cite{mcpdft,mcpdft_ct,mcpdft_diff,mcpdft_exc,odoh_2016,ghosh_2017,Sand2017,liam_2017,MCPDFT_ACR}.

In this work, we present the formulation and first application of analytic gradients for MC-PDFT, using state-specific~(SS-) CASSCF wave functions as references~(SS-CAS-PDFT).  However, SS-CAS-PDFT is not variational; the orbitals and CI expansion coefficients, while variationally optimized to minimize the energy for the reference CASSCF calculation, are not variationally optimized at the SS-CAS-PDFT level.  This means the Helmann-Feynman theorem~\cite{Hellmann,Feynman,Kern,Nakatsuji} does not hold, and coupled-perturbed MCSCF equations~\cite{hoffmann_1984,helgaker_1986,Page_1984,Grad_book,dudley_2006} must be solved.
However, because the SS-CASSCF reference wave function is variational, it is possible to construct a Lagrangian that is variational~\cite{helgaker_1989,Bernhardsson,Stalring}.  This allows us to circumvent the use of coupled-perturbed MCSCF.

Our paper is organized as follows. First, we provide a brief background on MC-PDFT theory.  We next provide the derivation of the equations used to compute MC-PDFT gradients analytically.  Finally, we demonstrate the use of these equations by performing benchmark geometry optimizations on ammonia and a set of nine small- to medium-sized organic molecules.

\section{Theory} \label{sec:theory}
We begin by giving a brief background on MC-PDFT theory, then we discuss the use of Lagrange multipliers to address the non-variationality of the SS-CAS-PDFT energy, and finally we present the analytic form of the MC-PDFT gradient.  Throughout this section we use the indices $p,q,r,s,t,...$ to refer to general molecular orbitals.  We also make use of the following excitation operators:
\begin{align}
\hat{E}_{pq} &= \hat{a}_{p\alpha}^\dag \hat{a}_{q\alpha} + \hat{a}_{p\beta}^\dag \hat{a}_{q\beta} \\
\hat{E}_{pq}^- &= \hat{E}_{pq}-\hat{E}_{qp} \\
\hat{e}_{pqrs} &= \hat{E}_{pq}\hat{E}_{rs} - \delta_{qr}\hat{E}_{ps}
\end{align}
where $\hat{a}_{p\alpha}^{\dag}$ and $\hat{a}_{p\alpha}$ are second-quantized creation and annihilation operators, respectively, and $\alpha$ and $\beta$ are spin quantum numbers.

\subsection{MC-PDFT theory}
The energy in MC-PDFT can be expressed as~\cite{mcpdft,Sand2017}
\begin{equation}
E_{\mathrm{PDFT}} = V_\mathrm{nn} + \sum_{pq}h_{pq}D_{pq} + \frac{1}{2}\sum_{pqst}g_{pqst}D_{pq}D_{st} + E_{\mathrm{ot}}[\rho,\Pi,\rho',\Pi']
\end{equation}
where $h_{pq}$ and $g_{pqst}$ are the one- and two-electron integrals:
\begin{equation}
h_{pq} = \int \phi_p^*({\bf{r}})h({\bf{r}})\phi_q({\bf{r}})d{\bf{r}}
\end{equation}
\begin{equation}
g_{pqst} = \int \int \phi_s^*({\bf{r}}_1)\phi_t({\bf{r}}_1)\frac{1}{|{\bf{r}}_1 - {\bf{r}}_2|} \phi_p^*({\bf{r}}_2)\phi_q({\bf{r}}_2)d{\bf{r}}_1d{\bf{r}}_2
\end{equation}
where the one-electron operator $h({\bf{r}})$ accounts for both electronic kinetic energy and electron-nuclear potential energy:
\begin{equation}
h({\bf{r}}) = \frac{-\nabla^2}{2} - \sum_A \frac{Z_A}{|{\bf{r}} - {\bf{r}}_A|},
\end{equation}
Molecular orbitals are indicated by $\phi_i$  (we assume real orbitals to keep the notation simple), while $\rho$ and $\Pi$ are the electronic density and the on-top pair density, respectively, and $\rho'$ and $\Pi'$ are their derivatives.  These densities can be expressed in terms of the orbitals, the one-body density matrix $D$, and the two-body density matrix $d$:
\begin{equation}
\rho({\bf{r}}) = \sum_{pq}\phi_p({\bf{r}})\phi_q({\bf{r}})D_{pq}
\label{eq:rho_def}
\end{equation}
\begin{equation}
\Pi({\bf{r}}) = \sum_{pqst}\phi_p({\bf{r}})\phi_q({\bf{r}})\phi_s({\bf{r}})\phi_t({\bf{r}})d_{pqst}
\label{eq:pi_def}
\end{equation}
\begin{equation}
\rho'({\bf{r}}) = \sum_{pq}\big{(} \phi'_p({\bf{r}})\phi_q({\bf{r}}) + \phi_p({\bf{r}})\phi_q'({\bf{r}}) \big{)}D_{pq}
\end{equation}
\begin{multline}
\Pi'({\bf{r}}) = \sum_{pqst} \big{(} \phi'_p({\bf{r}})\phi_q({\bf{r}})\phi_s({\bf{r}})\phi_t({\bf{r}}) +\phi_p({\bf{r}})\phi'_q({\bf{r}})\phi_s({\bf{r}})\phi_t({\bf{r}}) \\ + \phi_p({\bf{r}})\phi_q({\bf{r}})\phi'_s({\bf{r}})\phi_t({\bf{r}}) + \phi_p({\bf{r}})\phi_q({\bf{r}})\phi_s({\bf{r}})\phi'_t({\bf{r}})  \big{)} d_{pqst}
\end{multline}
The current generation of on-top energy functionals $E_{\mathrm{ot}}[\rho,\Pi,\rho',\Pi']$ is formed by translating existing local density approximation and generalized gradient approximation Kohn-Sham~(KS) density functionals~\cite{mcpdft,mcpdft_ft}.  The parent KS density functionals depend on the spin-up and spin-down electron densities $\rho_{\alpha}$ and $\rho_{\beta}$ as well as their derivatives $\rho_{\alpha}'$ and $\rho_{\beta}'$.  We express these KS functionals as $E_{\mathrm{xc}}[\rho_{\alpha},\rho_{\beta},\rho_{\alpha}',\rho_{\beta}']$.  Our original translation scheme~\cite{mcpdft} defines the on-top energy functional (with no dependence on $\Pi'$) as
\begin{equation}
E_{\mathrm{ot}}[\rho,\Pi,\rho'] = E_{\mathrm{xc}}[\tilde{\rho}_{\alpha},\tilde{\rho}_{\beta},\tilde{\rho}_{\alpha}',\tilde{\rho}_{\beta}']
\end{equation}
where the translated densities are defined as
\begin{align}
 \tilde{\rho}_{\alpha}({\bf{r}}) &= 
 \begin{cases}
 \frac{\rho({\bf{r}})}{2} (1+\zeta_t(\bf{r})) &  R({\bf{r}}) \leq 1 \\
 \frac{\rho({\bf{r}})}{2}  & R({\bf{r}}) > 1
 \end{cases} \label{eq:trans1} \\
 \tilde{\rho}_{\beta}({\bf{r}}) &= 
  \begin{cases}
 \frac{\rho({\bf{r}})}{2} (1-\zeta_t(\bf{r})) &  R({\bf{r}}) \leq 1 \\
 \frac{\rho({\bf{r}})}{2}  & R({\bf{r}}) > 1
 \end{cases} \\
 \tilde{\rho}_{\alpha}'({\bf{r}}) &= 
  \begin{cases}
 \frac{\rho'({\bf{r}})}{2} (1+\zeta_t(\bf{r})) &  R({\bf{r}}) \leq 1 \\
 \frac{\rho'({\bf{r}})}{2}  & R({\bf{r}}) > 1
 \end{cases} \\
 \tilde{\rho}_{\beta}'({\bf{r}}) &= 
  \begin{cases}
 \frac{\rho'({\bf{r}})}{2} (1-\zeta_t(\bf{r})) &  R({\bf{r}}) \leq 1 \\
 \frac{\rho'({\bf{r}})}{2}  & R({\bf{r}}) > 1
 \end{cases} \label{eq:trans4}
\end{align}
where
\begin{equation}
\zeta_t(\bf{r}) = \sqrt{1-R(\\{\bf{r}})}
\end{equation}
\begin{equation}
R({\bf{r}}) = \frac{\Pi({\bf{r}})}{[\rho({\bf{r}})/2]^2}
\end{equation}
and
\begin{align}
\rho({\bf{r}}) = \rho_{\alpha}({\bf{r}}) + \rho_{\beta}({\bf{r}}) \\
\rho'({\bf{r}}) = \rho_{\alpha}'({\bf{r}}) + \rho_{\beta}'({\bf{r}}).
\end{align}
This scheme is denoted by ``t''.  We have also developed a fully-translated (``ft'') scheme~\cite{mcpdft_ft} in which the on-top functional depends on $\Pi'$.  For simplicity, the main text will only treat the ``t'' functional case, but we present an appendix that gives the changes for the ``ft'' functional case.

\subsection{SS-CAS-PDFT energy Lagrangian}
The Hellmann-Feynman theorem~\cite{Hellmann,Feynman,Kern,Nakatsuji} shows that if a wave function is variationally optimized in all of its parameters, the first-order energy response to a perturbation depends only on the expectation value of the derivative of the Hamiltonian operator with respect to the perturbation.
\begin{equation}
\frac{d}{d \lambda} \langle \Psi_{\mathrm{var}} | \hat{H} | \Psi_{\mathrm{var}} \rangle = \langle \Psi_{\mathrm{var}} | \frac{d\hat{H}}{d\lambda} | \Psi_{\mathrm{var}} \rangle
\end{equation}
In this work, we will only use SS-CASSCF wave functions as references for MC-PDFT calculations. In a SS-CASSCF wave function, the energy is parameterized in the following way~\cite{bible}:
\begin{equation}
E_0 = \langle 0 | e^{\hat{P}} e^{\hat{\kappa}} \hat{H} e^{-\hat{\kappa}}e^{-\hat{P}} | 0 \rangle
\end{equation} 
where $\hat{\kappa}$ represents the orbital rotation operators
\begin{equation}
\hat{\kappa} = \sum_{p<q} \kappa_{pq}\hat{E}_{pq}^-
\end{equation}
where $p$ and $q$ are orbital indices, and $\hat{P}$ represents the CI state transfer operators
\begin{equation}
\hat{P} = \sum_I P_I(|I\rangle \langle 0| - |0\rangle \langle I|)
\end{equation}
where $|I\rangle$ corresponds to a configuration state function~(CSF).  In a SS-CASSCF calculation, one variationally optimizes the non-redundant values of $\kappa_{pq}$ and $P_I$ to minimize the energy within a (non-variational) finite basis set.  Because the SS-CASSCF wave function is variational in all wave function parameters, only the derivative of the Hamiltonian and derivatives of the basis set parameters can contribute to the SS-CASSCF gradient.  In the case of SS-CAS-PDFT, however, the final energy is not variational in either the wave function parameters (the CI coefficients and the orbital coefficients) or the basis set parameters.  Therefore, derivatives of the wave function provide nonzero contributions to the energy gradient.  In general, the solutions to coupled-perturbed MCSCF equations would be needed.  For a SS-CASSCF wave function, however, it is possible to use Lagrange multiplier techniques~\cite{helgaker_1989,Bernhardsson,Stalring} to write an expression for a Lagrangian which is variational in all wave function degrees of freedom.
The SS-CAS-PDFT Lagrangian is given by
\begin{equation}
L(\textbf{c},\textbf{z}) = E_{\mathrm{PDFT}} + \sum_{pq} z_{pq} \frac{\partial E_0}{\partial \kappa_{pq}} + \sum_{I} z_I \frac{\partial E_0}{\partial P_I}
\label{eq:lagrangian}
\end{equation}
\begin{equation}
\frac{\partial L}{\partial \lambda} = \frac{\partial E_{\mathrm{PDFT}}}{\partial \lambda}
\end{equation}
where $E_0$ is the SS-CASSCF state energy, $z_{pq}$ is a Lagrange multiplier associated with orbital rotation, and $z_I$ is a Lagrange multiplier associated with a CI state transfer.  The values of $z_{pq}$ and $z_I$ are determined by taking the derivative of eq.~\ref{eq:lagrangian} with respect to each wave function parameter and setting the resultant expressions to zero.
These derivatives are written as
\begin{align}
\frac{\partial L}{\partial \kappa_{pq}} &= \frac{\partial E_{\mathrm{PDFT}}}{\partial \kappa_{pq}} + \sum_{st} z_{st} \frac{\partial^2 E_0}{\partial \kappa_{st} \partial \kappa_{pq}} + \sum_J z_J \frac{\partial^2 E_0}{\partial P_J \partial \kappa_{pq}} = 0 \label{eq:zqn1} \\
\frac{\partial L}{\partial P_{I}} &= \frac{\partial E_{\mathrm{PDFT}}}{\partial P_{I}} + \sum_{st} z_{st} \frac{\partial^2 E_0}{\partial \kappa_{st} \partial P_{I}} + \sum_J z_J \frac{\partial^2 E_0}{\partial P_J \partial P_{I}} = 0 \label{eq:zqn2}
\end{align}
Thus, the determination of $z_{pq}$ and $z_I$ requires the SS-CASSCF Hessian matrix and the first-order response to the PDFT energy from changes in $\kappa_{xy}$ and $P_I$.  Eqs.~\ref{eq:zqn1}--\ref{eq:zqn2} generate a set of linear equations which can be written in matrix form as
\begin{equation}
\begin{pmatrix}
A_{\kappa \kappa} & A_{\kappa P} \\
A_{P \kappa} & A_{PP}
\end{pmatrix}
\begin{pmatrix}
z_{\kappa} \\
z_{P}
\end{pmatrix}
=
\begin{pmatrix}
\partial E_{\mathrm{PDFT}} / \partial \kappa \\
\partial E_{\mathrm{PDFT}} / \partial P
\end{pmatrix}
\label{eq:mat_lag}
\end{equation}
where $\bf A$ is the SS-CASSCF Hessian with elements defined by
\begin{align}
A_{\kappa_{st} \kappa_{pq}} &= \frac{\partial^2 E_0}{\partial \kappa_{st} \partial \kappa_{pq}} \\
A_{P_J \kappa_{pq}} &= \frac{\partial^2 E_0}{\partial P_J \partial \kappa_{pq}} \\
A_{P_I \kappa_{st}} &= \frac{\partial^2 E_0}{\partial P_I \partial \kappa_{st}} \\
A_{P_I P_J} &= \frac{\partial^2 E_0}{\partial P_J \partial P_{I}}
\end{align}

\subsection{SS-CAS-PDFT orbital and CI responses}
The right-hand side of eq.~\ref{eq:mat_lag} requires the derivative of the SS-CAS-PDFT energy with respect to both orbital rotation parameters and CI coefficients.  First, we examine the orbital rotation response:
\begin{equation}
\frac{\partial E_{\mathrm{PDFT}}}{\partial \kappa_{xy}} = \frac{\partial}{\partial \kappa_{xy}} \Big{(} V_{\mathrm{nn}} + \sum_{pq}h_{pq} D_{pq} + \frac{1}{2}\sum_{pqst}g_{pqst}D_{pq}D_{st} + E_{\mathrm{ot}}[\rho,\Pi,\rho']\Big{)}. \label{eq:orbgrad}
\end{equation}
While $V_{\mathrm{nn}}, h_{pq},$ and $g_{pqst}$ are independent of $\kappa_{xy}$, the one-body density matrix and the on-top energy functional are not, and these terms will contribute to the derivative in a non-zero fashion.  The derivatives of $D$ and $d$ with respect to $\kappa_{xy}$, through first-order, are
\begin{equation}
\frac{\partial D_{pq}}{\partial \kappa_{xy}} = \delta_{yp}D_{xq} - \delta_{xp} D_{yq} - \delta_{xq}D_{py} + \delta_{yq}D_{px}
\label{eq:1D_kap}
\end{equation}
\begin{equation}
\frac{\partial d_{pqst}}{\partial \kappa_{xy}} = \delta_{yp}d_{xqst} - \delta_{xp}d_{yqst} + \delta_{yq}d_{pxst} - \delta_{xq}d_{pyst} + \delta_{ys}d_{pqxt} - \delta_{xs}d_{pqyt} + \delta_{yt}d_{pqsx} + \delta_{xt}d_{pqsy},
\label{eq:2d_kap}
\end{equation}
where $\delta_{pq}$ is the Kronecker delta.
To calculate the derivative of the on-top energy functional, we can use the definitions of $\rho({\bf r})$~(eq.~\ref{eq:rho_def}) and $\Pi({\bf r})$~(eq.~\ref{eq:pi_def}) and the chain rule to obtain
\begin{equation}
\frac{\partial E_\mathrm{ot}[\rho,\Pi,\rho']}{\partial \kappa_{xy}} = \sum_{pq} \frac{\partial E_\mathrm{ot}[\rho,\Pi,\rho']}{\partial D_{pq}} \frac{\partial D_{pq}}{\partial \kappa_{xy}} + \sum_{pqst}\frac{\partial E_\mathrm{ot}[\rho,\Pi,\rho']}{\partial d_{pqst}} \frac{\partial d_{pqst}}{\partial \kappa_{xy}}. 
\end{equation}
We define the one-electron on-top potential as
\begin{equation}
V_{pq} = \frac{\partial E_\mathrm{ot}[\rho,\Pi,\rho']}{\partial D_{pq}} \label{eq:oeot}.
\end{equation}
By using the translation scheme in eqs.~(\ref{eq:trans1}-\ref{eq:trans4}), we can write the one-electron on-top potential in terms of the derivatives of the KS density functional (which depends on the translated densities and derivatives denoted by tildes):
\begin{multline}
V_{pq} =
\frac{\partial E_{\mathrm{xc}}[\tilde{\rho}_{\alpha},\tilde{\rho}_{\beta},\tilde{\rho}_{\alpha}',\tilde{\rho}_{\beta}']}{\partial \tilde{\rho}_{\alpha}} \frac{\partial \tilde{\rho}_{\alpha}}{\partial D_{pq}} +
\frac{\partial E_{\mathrm{xc}}[\tilde{\rho}_{\alpha},\tilde{\rho}_{\beta},\tilde{\rho}_{\alpha}',\tilde{\rho}_{\beta}']}{\partial \tilde{\rho}_{\beta}} \frac{\partial \tilde{\rho}_{\beta}}{\partial D_{pq}} \\ +
\frac{\partial E_{\mathrm{xc}}[\tilde{\rho}_{\alpha},\tilde{\rho}_{\beta},\tilde{\rho}_{\alpha}',\tilde{\rho}_{\beta}']}{\partial \tilde{\rho}_{\alpha}'} \frac{\partial \tilde{\rho}_{\alpha}'}{\partial D_{pq}} + 
\frac{\partial E_{\mathrm{xc}}[\tilde{\rho}_{\alpha},\tilde{\rho}_{\beta},\tilde{\rho}_{\alpha}',\tilde{\rho}_{\beta}']}{\partial \tilde{\rho}_{\beta}'} \frac{\partial \tilde{\rho}_{\beta}'}{\partial D_{pq}} \label{eq:pot1.2}
\end{multline}
The first component of each term in eq.~\ref{eq:pot1.2}, the KS density functional derivative, can be calculated numerically over a grid of points using existing DFT routines~\cite{perdew_1996}.  The second component of each term, the derivatives of the translated densities and gradients with respect to the one-body density matrix, are obtained by taking the derivative of eqs.~(\ref{eq:trans1}-\ref{eq:trans4}) over a grid of points:
\begin{align}
 \frac{\partial \tilde{\rho}_{\alpha}({\bf r})}{\partial D_{pq}} &= 
 \begin{cases}
 \bigg(\frac{1}{2}\Big[1+\zeta_t({\bf r})\Big]+\frac{2\Pi({\bf r})}{\zeta_t({\bf r})[\rho({\bf r})]^2}\bigg)\Big[\phi_p({\bf r})\phi_q({\bf r})\Big] &  R({\bf{r}}) \leq 1 \\
 \frac{1}{2}\Big[\phi_p({\bf r})\phi_q({\bf r})\Big] & R({\bf{r}}) > 1
 \end{cases}  \\
 \frac{\partial \tilde{\rho}_{\beta}({\bf r})}{\partial D_{pq}} &= 
  \begin{cases}
 \bigg(\frac{1}{2}\Big[1-\zeta_t({\bf r})\Big]-\frac{2\Pi({\bf r})}{\zeta_t({\bf r})[\rho({\bf r})]^2}\bigg)\Big[\phi_p({\bf r})\phi_q({\bf r})\Big] &  R({\bf{r}}) \leq 1 \\
 \frac{1}{2}\Big[\phi_p({\bf r})\phi_q({\bf r})\Big] & R({\bf{r}}) > 1
 \end{cases} \\
 \frac{\partial \tilde{\rho}_{\alpha}'({\bf r})}{\partial D_{pq}} &= 
  \begin{cases}
 \frac{1}{2}\Big[1+\zeta_t({\bf r})\Big]\Big[\phi_p'({\bf r})\phi_q({\bf r}) + \phi_p({\bf r})\phi_q'({\bf r})\Big] + \frac{2\rho'({\bf r})\Pi({\bf r})}{\zeta_t({\bf r})[\rho({\bf r})]^3}\Big[\phi_p({\bf r})\phi_q({\bf r})\Big] & R({\bf{r}}) \leq 1 \\
 \frac{1}{2}\Big[\phi_p'({\bf r})\phi_q({\bf r}) + \phi_p({\bf r})\phi_q'({\bf r})\Big] & R({\bf{r}}) > 1
 \end{cases} \\
 \frac{\partial \tilde{\rho}_{\beta}'({\bf r})}{\partial D_{pq}} &= 
  \begin{cases}
  \frac{1}{2}\Big[1-\zeta_t({\bf r})\Big]\Big[\phi_p'({\bf r})\phi_q({\bf r}) + \phi_p({\bf r})\phi_q'({\bf r})\Big] - \frac{2\rho'({\bf r})\Pi({\bf r})}{\zeta_t({\bf r})[\rho({\bf r})]^3}\Big[\phi_p({\bf r})\phi_q({\bf r})\Big] & R({\bf{r}}) \leq 1 \\
 \frac{1}{2}\Big[\phi_p'({\bf r})\phi_q({\bf r}) + \phi_p({\bf r})\phi_q'({\bf r})\Big] & R({\bf{r}}) > 1
 \end{cases} 
\end{align}
Likewise, we define the two-electron on-top potential as 
\begin{equation}
v_{pqst} = \frac{\partial E_\mathrm{ot}[\rho,\Pi,\rho']}{\partial d_{pqst}}. \label{eq:teot}
\end{equation}
Similarly, by using the translation scheme in eqs.~(\ref{eq:trans1}-\ref{eq:trans4}), we can write the two-electron on-top potential in terms of the derivatives of the KS density functional:
\begin{multline}
v_{pqst} =
\frac{\partial E_{\mathrm{xc}}[\tilde{\rho}_{\alpha},\tilde{\rho}_{\beta},\tilde{\rho}_{\alpha}',\tilde{\rho}_{\beta}']}{\partial \tilde{\rho}_{\alpha}} \frac{\partial \tilde{\rho}_{\alpha}}{\partial d_{pqst}} +
\frac{\partial E_{\mathrm{xc}}[\tilde{\rho}_{\alpha},\tilde{\rho}_{\beta},\tilde{\rho}_{\alpha}',\tilde{\rho}_{\beta}']}{\partial \tilde{\rho}_{\beta}} \frac{\partial \tilde{\rho}_{\beta}}{\partial d_{pqst}} \\ +
\frac{\partial E_{\mathrm{xc}}[\tilde{\rho}_{\alpha},\tilde{\rho}_{\beta},\tilde{\rho}_{\alpha}',\tilde{\rho}_{\beta}']}{\partial \tilde{\rho}_{\alpha}'} \frac{\partial \tilde{\rho}_{\alpha}'}{\partial d_{pqst}} + 
\frac{\partial E_{\mathrm{xc}}[\tilde{\rho}_{\alpha},\tilde{\rho}_{\beta},\tilde{\rho}_{\alpha}',\tilde{\rho}_{\beta}']}{\partial \tilde{\rho}_{\beta}'} \frac{\partial \tilde{\rho}_{\beta}'}{\partial d_{pqst}}
\end{multline}
The derivatives of the translated densities and gradients with respect to the two-body density matrix are obtained by taking derivatives of eqs.~(\ref{eq:trans1}-\ref{eq:trans4}) over a grid of points:
\begin{align}
 \frac{\partial \tilde{\rho}_{\alpha}({\bf r})}{\partial d_{pqst}} &= 
 \begin{cases}
 \bigg(\frac{-1}{\rho({\bf r})\zeta_t({\bf r})} \bigg)\Big[\phi_p({\bf r})\phi_q({\bf r})\phi_s({\bf r})\phi_t({\bf r})\Big] &  R({\bf{r}}) \leq 1 \\
 0 & R({\bf{r}}) > 1
 \end{cases}  \\
 \frac{\partial \tilde{\rho}_{\beta}({\bf r})}{\partial d_{pqst}} &= 
  \begin{cases}
\bigg(\frac{1}{\rho({\bf r})\zeta_t({\bf r})} \bigg)\Big[\phi_p({\bf r})\phi_q({\bf r})\phi_s({\bf r})\phi_t({\bf r})\Big] &  R({\bf{r}}) \leq 1 \\
 0 & R({\bf{r}}) > 1
 \end{cases} \\
 \frac{\partial \tilde{\rho}_{\alpha}'({\bf r})}{\partial d_{pqst}} &= 
 \begin{cases}
 \bigg( \frac{- \rho'({\bf r})}{[\rho({\bf r})]^2\zeta_t({\bf r})} \bigg) \Big[ \phi_p({\bf r})\phi_q({\bf r})\phi_s({\bf r})\phi_t({\bf r}) \Big]  & R({\bf{r}}) \leq 1 \\
0 & R({\bf{r}}) > 1
 \end{cases} \\
 \frac{\partial \tilde{\rho}_{\beta}'({\bf r})}{\partial d_{pqst}} &= 
  \begin{cases}
  \bigg(\frac{\rho'({\bf r})}{[\rho({\bf r})]^2\zeta_t({\bf r})} \bigg)\Big[\phi_p({\bf r})\phi_q({\bf r})\phi_s({\bf r})\phi_t({\bf r})\Big] & R({\bf{r}}) \leq 1 \\
 0 & R({\bf{r}}) > 1
 \end{cases} 
\end{align}
By combining these results, eq.~\ref{eq:orbgrad} can be written succinctly as
\begin{equation}
\frac{\partial E_{\mathrm{PDFT}}}{\partial \kappa_{xy}} = 2(F_{xy} - F_{yx})
\end{equation}
where $F_{xy}$ are elements of the generalized SS-CAS-PDFT Fock matrix:
\begin{equation}
F_{xy} = \sum_p (h_{py} + V_{py})D_{px} + \sum_{pqs}(g_{pqry}D_{pq}D_{rx} + 2v_{pqry}d_{pqrx}).
\end{equation} 

Next, we consider the CI state transfer response:
\begin{equation}
\frac{\partial E_{\mathrm{PDFT}}}{\partial P_I} = \frac{\partial}{\partial P_I} \Big{(} V_{\mathrm{nn}} + \sum_{pq}h_{pq} D_{pq} + \frac{1}{2}\sum_{pqst}g_{pqst}D_{pq}D_{st} + E_\mathrm{ot}[\rho,\Pi,\rho']\Big{)}. \label{eq:cigrad}
\end{equation}
The derivative of the one-body and two-body density matrix with respect to a CI coefficient are given by
\begin{equation}
\frac{\partial D_{pq}}{\partial P_i} = \langle i | \hat{E}_{pq} | 0 \rangle + \langle 0 | \hat{E}_{pq} |i \rangle - 2 \langle i |0 \rangle D_{pq}
\label{eq:1d_ci}
\end{equation}
\begin{equation}
\frac{\partial d_{pqst}}{\partial P_i} = \langle i | \hat{e}_{pqst} | 0 \rangle + \langle 0 | \hat{e}_{pqst} |i \rangle - 2 \langle i |0 \rangle d_{pqst}
\label{eq:2d_ci}
\end{equation}
where $| 0 \rangle$ is the SS-CASSCF reference wavefunction and $| i \rangle$ is a CSF.
The derivative of the on-top energy functional with respect to a CI coefficient is given as
\begin{align}
\frac{\partial E_\mathrm{ot}[\rho,\Pi,\rho']}{\partial P_i} &= \sum_{pq} \frac{\partial E_\mathrm{ot}[\rho,\Pi,\rho']}{\partial D_{pq}} \frac{\partial D_{pq}}{\partial P_i} + \sum_{pqst}\frac{\partial E_\mathrm{ot}[\rho,\Pi,\rho']}{\partial d_{pqst}} \frac{\partial d_{pqst}}{\partial P_i} \\
&= \sum_{pq} V_{pq} \frac{\partial D_{pq}}{\partial P_i} + \sum_{pqst} v_{pqst} \frac{\partial d_{pqst}}{\partial P_i}
\end{align}
These results allow us to write eq.~\ref{eq:cigrad} as
\begin{equation}
\frac{\partial E_{\mathrm{PDFT}}}{\partial P_i} = 2 \langle i | \hat{F}_1 + \hat{F}_2 - E | 0 \rangle
\end{equation}
where
\begin{equation}
\hat{F}_1 = \sum_{pq} \bigg[ h_{pq} + V_{pq} + \sum_{st} g_{pqst}D_{rs} \bigg] \hat{E}_{pq}
\end{equation}
\begin{equation}
\hat{F}_2 = \sum_{pqst} v_{pqst} \hat{e}_{pqst}
\end{equation}
\begin{equation}
E= \langle 0 | \hat{F}_1 + \hat{F}_2 | 0 \rangle .
\end{equation}

\subsection{Response equations}
The Lagrange multipliers are obtained by solving the linear system of equations given in eq.~\ref{eq:mat_lag}.  In lieu of the explicit construction and direct diagonalization of the SS-CASSCF Hessian matrix, we solve eq.~\ref{eq:mat_lag} via the iterative preconditioned conjugate gradient~(PCG) algorithm~\cite{Fortran}, during which Hessian and trial vector multiplications are performed on-the-fly.  In our approach, the standard CASSCF preconditioner is used~\cite{Bernhardsson}.  Further details about the PCG procedure are available in refs.~\cite{Bernhardsson,Stalring}

\subsection{SS-CAS-PDFT nuclear gradients}
The gradient of the SS-CAS-PDFT energy, upon determination of the Lagrange multipliers, is given by
\begin{equation}
\frac{d E_\mathrm{{PDFT}}}{d \lambda} = E_{\mathrm{PDFT}}^{(\lambda)} + \sum_{xy} z_{xy} \frac{\partial}{\partial \kappa_{xy}} (\sum_{pq} h_{pq}^{\lambda}D_{pq} + \sum_{pqst} g_{pqst}^{\lambda}d_{pqst}) +   \sum_{I} z_{I} \frac{\partial}{\partial P_{I}} (\sum_{pq} h_{pq}^{\lambda}D_{pq} + \sum_{pqst} g_{pqst}^{\lambda}d_{pqst})
\label{eq:fingrad}
\end{equation}
where $h_{pq}^{\lambda}$ and $g_{pqst}^{\lambda}$ are derivative integrals given by
\begin{align}
h_{pq}^{\lambda} &= \frac{d h_{pq}}{d \lambda} \\
&= \frac{\partial h_{pq}}{\partial \lambda} - \frac{1}{2}\sum_w \Big[ \frac{\partial S_{pw}}{\partial \lambda}h_{wq} +  \frac{\partial S_{wq}}{\partial \lambda}h_{pw} \Big] \\
g_{pqst}^{\lambda} &= \frac{d g_{pqst}}{d \lambda} \\
&= \frac{\partial g_{pqst}}{\partial \lambda} - \frac{1}{2}\sum_w \Big[ \frac{\partial S_{pw}}{\partial \lambda}g_{wqst} + \frac{\partial S_{qw}}{\partial \lambda}g_{pwst} +  \frac{\partial S_{sw}}{\partial \lambda}g_{pqwt} + \frac{\partial S_{tw}}{\partial \lambda}g_{pqsw}  \Big]
\end{align}
where $S_{xy}$ is an element of the orbital overlap matrix.  The terms involving derivatives of the overlap matrix arise due to the response of the basis set to the perturbation.  This is often referred to as the `connection' or `renormalization' contribution.  $E_{\mathrm{PDFT}}^{(\lambda)}$ is the SS-CAS-PDFT energy expression evaluated with the derivatives of energy operators and functionals:
\begin{equation}
E_{\mathrm{PDFT}}^{(\lambda)} = \frac{d V_\mathrm{nn}}{d \lambda} + \sum_{pq} h_{pq}^{\lambda} D_{pq} + \frac{1}{2}\sum_{pqst} g_{pqst}^{\lambda} D_{pq}D_{st} + \frac{d E_{\mathrm{ot}}[\rho,\Pi,\rho']}{d \lambda}
\end{equation}
Like the one- and two-electron integral derivatives, the derivative of the on-top functional also contributes a `renormalization' contribution: 
\begin{equation}
 \frac{d E_{\mathrm{ot}}[\rho,\Pi,\rho']}{d \lambda} = \sum_{xy} S_{xy}^{\lambda} \Big[  \sum_p V_{xp}D_{yp} + \sum_{pqs} v_{xpqs}d_{ypqs} \Big] + \frac{\partial E_{\mathrm{ot}}[\rho,\Pi,\rho']}{\partial \lambda}.\label{eq:dftg}
\end{equation}
The evaluation of the final term in eq.~\ref{eq:dftg} is evaluated using standard DFT techniques~\cite{}, using the translated densities in the evaluation of the derivative of the KS-DFT functional.
The remaining derivatives in eq.~\ref{eq:fingrad} can be evaluated using eqs.~\ref{eq:1D_kap}-\ref{eq:2d_kap} and eqs.~\ref{eq:1d_ci}-\ref{eq:2d_ci}, and upon rearrangement we obtain
\begin{equation}
\frac{d E_{\mathrm{PDFT}}}{d \lambda} = \frac{dV_{\mathrm{nn}}}{d\lambda} -\frac{1}{2}\sum_{pq} \frac{\partial S_{pq}}{\partial \lambda} F_{pq}^{\mathrm{eff}} + \sum_{pq}\frac{\partial h_{pq}}{\partial \lambda} D_{pq}^{\mathrm{eff}} + \sum_{pqst} \frac{\partial g_{pqst}}{\partial \lambda} d_{pqst}^{\mathrm{eff}} + \frac{\partial E_{\mathrm{ot}}[\rho,\Pi,\rho']}{\partial \lambda}
\end{equation}
where we have introduced an effective one-body density matrix
\begin{align}
D_{pq}^{\mathrm{eff}} &= D_{pq} + \breve{D}_{pq} + \bar{D}_{pq} \\
\breve{D}_{pq} &= \sum_{s}(D_{sq}z_{ps} - D_{ps}z_{sq}) \\
\bar{D}_{pq} &= \sum_I z_I(\langle I|\hat{E}_{pq}|0\rangle + \langle 0|\hat{E}_{pq}|I\rangle,
\end{align}
an effective two-body density matrix
\begin{align}
d_{pqst}^{\mathrm{eff}} &= d_{pqst} + \breve{d}_{pqst} + \bar{d}_{pqst} \\
\breve{d}_{pqst} &= \sum_t(d_{tqrs}z_{pt} - d_{ptrs}z_{qt} + d_{pqts}z_{rt} - d_{pqrt}z_{st}) \\
\bar{d}_{pqst} &= \sum_I z_I(\langle I|\hat{e}_{pqst}|0\rangle + \langle 0|\hat{e}_{pqst}|I\rangle,
\end{align}
and an effective Fock matrix
\begin{equation}
F_{pq}^{\mathrm{eff}} = \sum_t(h_{pt} + v_{pt})D_{qt} + \sum_{rst}g_{prst}D_{qr}D_{st} + \sum_{rst}v_{prst}d_{qrst} + \sum_t h_{pt}(\breve{D}_{qt} + \bar{D}_{qt}) + \sum_{rst}g_{prst}(\breve{d}_{qrst}+\bar{d}_{qrst}).
\end{equation}

\section{Methods}

\begin{figure}
\begin{center}
\includegraphics[scale=.30]{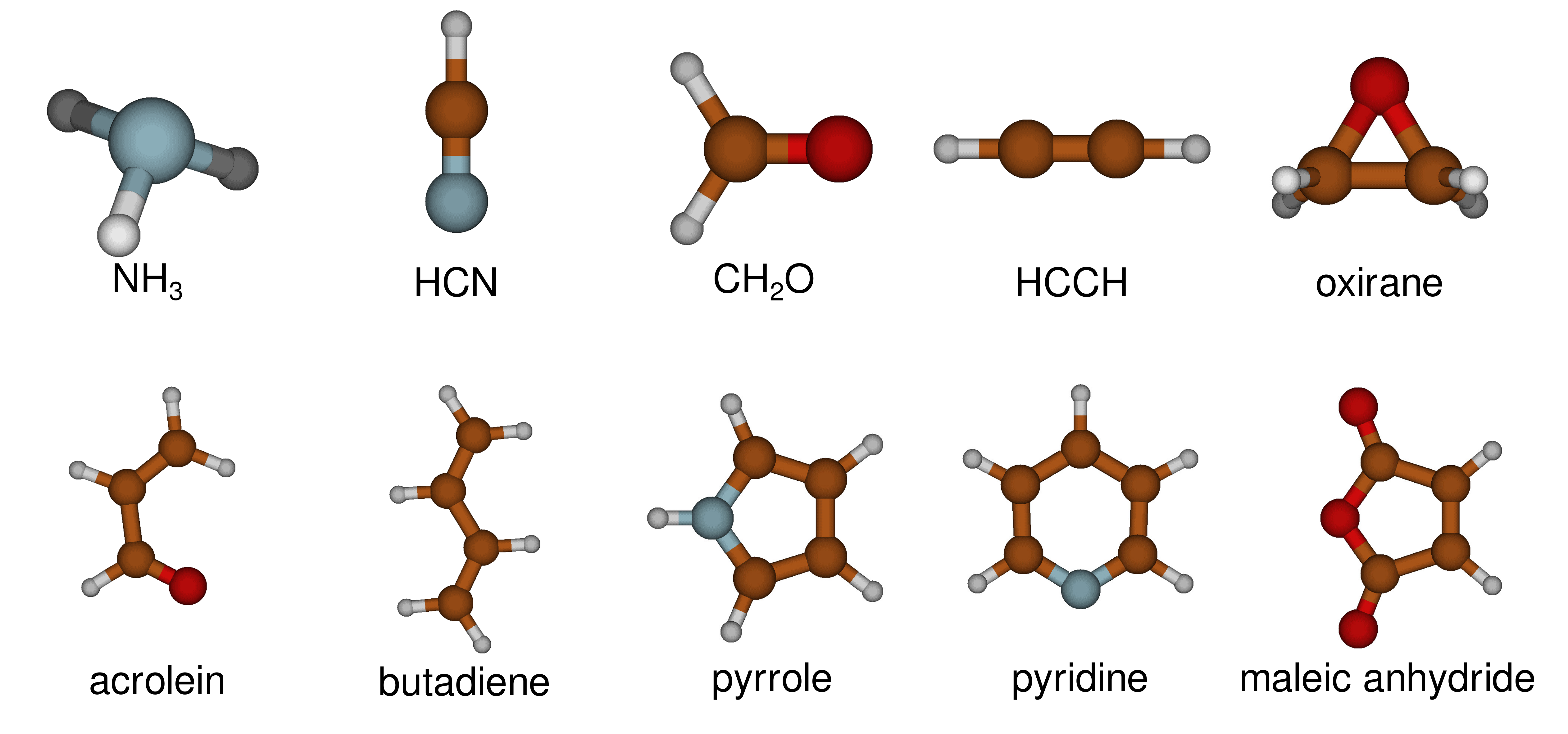}
\end{center}
\caption{Systems calculated in this study.}
\label{fig:systems}
\end{figure}

\begin{table}[htbp]
  \caption{Active space selection [denoted (electrons, orbitals)] and symmetry employed for each system.  The index is used in the labeling of Figs.~\ref{fig:ccd} and \ref{fig:cct}.}

    \begin{tabular}{lccc} \hline \hline
        & Index & Active Space  & Symmetry Constraints \\ \hline
    \textbf{NH$_3$} & 1 & (6,6) & C$_\mathrm{s}$ \\
    \textbf{HCN} & 2 & (8,8) & C$_\mathrm{2v}$ \\
    \textbf{CH$_2$O} & 3 & (12,9) & C$_\mathrm{2v}$ \\
    \textbf{HCCH} & 4 & (10,10) & D$_\mathrm{2h}$ \\
    \textbf{oxirane} & 5 & (10,10) & C$_\mathrm{2v}$ \\
    \textbf{acrolein} & 6 & (4,4) & C$_\mathrm{s}$ \\
    \textbf{butadiene} & 7 & (4,4) & C$_\mathrm{2h}$ \\
    \textbf{pyrrole} & 8 & (6,5) & C$_\mathrm{2v}$ \\
    \textbf{pyridine} & 9 & (6,6)  & C$_\mathrm{2v}$ \\
    \textbf{maleic anhydride} & 10 & (8,7) & C$_\mathrm{2v}$ \\ \hline
    \end{tabular}%
  \label{tab:param}%
\end{table}%

A set of ten molecules from the SE47 database of structures~\cite{Barone} was utilized in this study; these molecules are shown in Fig.~\ref{fig:systems}.  All CASSCF, CASPT2, Kohn-Sham density functional theory (KS-DFT), and MC-PDFT calculations were performed with a locally-modified version of \emph{Molcas 8.1}~\cite{Molcas8}.  The same CASSCF reference was used for all multireference calculations on a given system, and the active space selection and symmetry employed are listed in Table~\ref{tab:param}.  All MC-PDFT calculations employed the tPBE on-top density functional~\cite{mcpdft} and the fine integration grid.  An MC-PDFT numerical gradient implementation based on finite differences was also developed as part of this project.  All analytical results were verified to agree with the numerical results.  We employed the cc-pVDZ and cc-pVTZ basis sets~\cite{dunning1,dunning2} for all calculations. 

\section{Results}

\begin{figure}
\begin{center}
\includegraphics[scale=1]{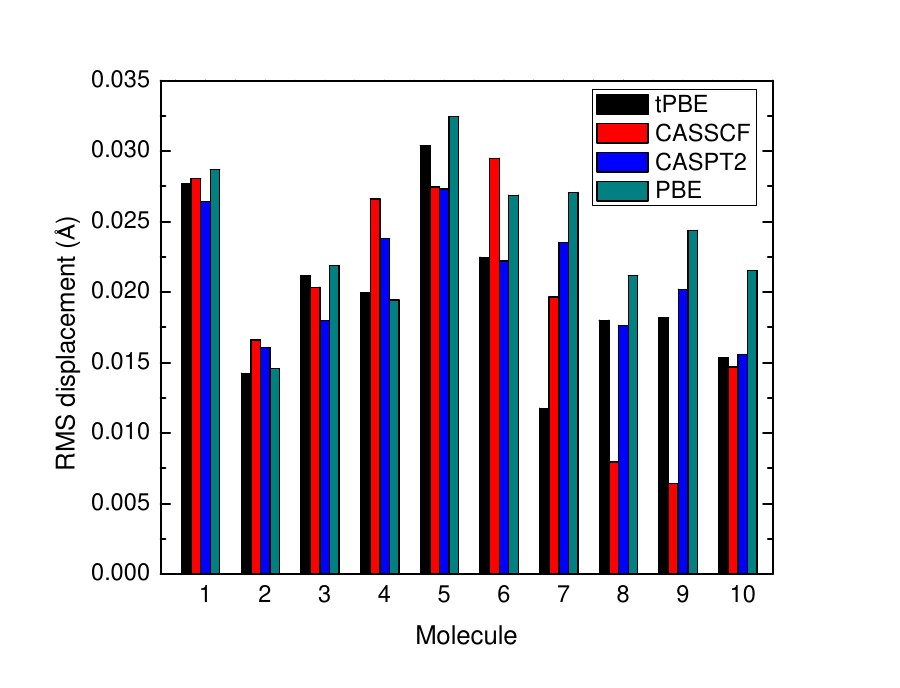}
\end{center}
\caption{Root mean square~(RMS) displacements of the atomic centers of the converged structure relative to the reference structure~\cite{Barone}.  The cc-pVDZ basis set was used.}
\label{fig:ccd}
\end{figure}

\begin{figure}
\begin{center}
\includegraphics[scale=1]{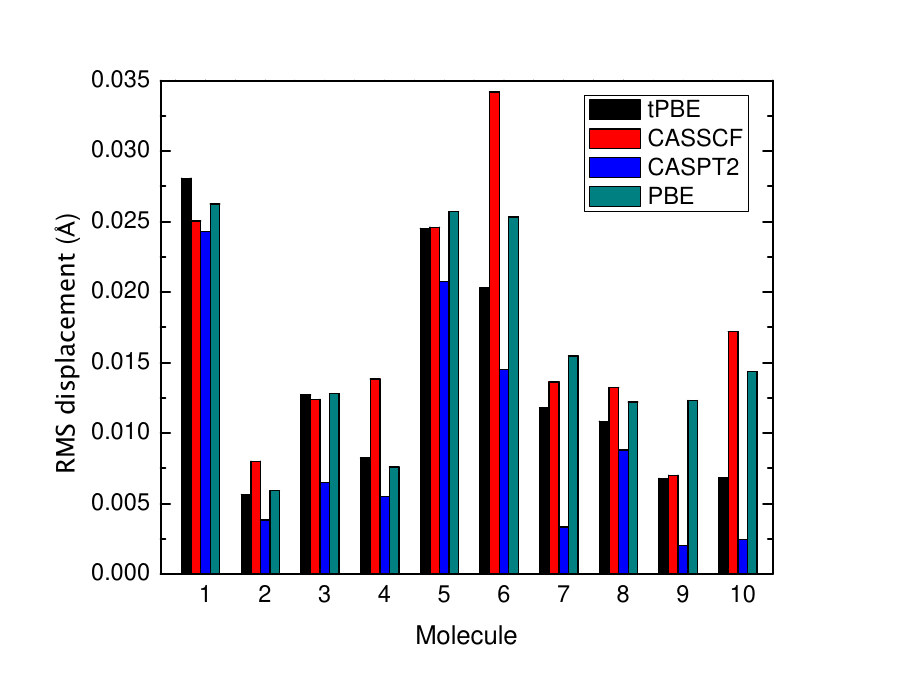}
\end{center}
\caption{RMS displacements of the atomic centers of the converged structure relative to the reference structure~\cite{Barone}.  The cc-pVTZ basis set was used.}
\label{fig:cct}
\end{figure}

\begin{table}[htbp]
  \caption{M-diagnostics~\cite{Mdiag} and orbital occupation numbers for the CASSCF equilibrium geometries.  M values greater than 0.1 indicate significant multireference character.}
    \begin{tabular}{lcccccc} \hline \hline
     & \multicolumn{2}{c}{M} & \multicolumn{2}{c}{HOMO occupation} & \multicolumn{2}{c}{LUMO occupation} \\  \cmidrule(lr){2-3} \cmidrule(lr){4-5} \cmidrule(lr){6-7}
          & cc-pVDZ  & cc-pVTZ & cc-pVDZ  & cc-pVTZ & cc-pVDZ  & cc-pVTZ\\ \hline
    \textbf{NH$_3$} & 0.023 & 0.022 & 1.977 & 1.978 & 0.022 & 0.023 \\
    \textbf{HCN} & 0.066 & 0.064 & 1.934 & 1.936 & 0.066 & 0.064 \\
    \textbf{CH$_2$O} & 0.065 & 0.063 & 1.936 & 1.939 & 0.067 & 0.064 \\
    \textbf{HCCH} & 0.068 & 0.065 & 1.932 & 1.935 & 0.068 & 0.066 \\
    \textbf{oxirane} & 0.036 & 0.034 & 1.965 & 1.966 & 0.036 & 0.035 \\
    \textbf{acrolein} & 0.107 & 0.103 & 1.897 & 1.900 & 0.110 & 0.106 \\
    \textbf{butadiene} & 0.118 & 0.113 & 1.884 & 1.889 & 0.120 & 0.116 \\
    \textbf{pyrrole} & 0.076 & 0.074 & 1.925 & 1.926 & 0.077 & 0.0753 \\
    \textbf{pyridine} &  0.103 & 0.101 & 1.898 & 1.900 & 0.105 & 0.102 \\
    \textbf{maleic anhydride} & 0.104 & 0.100 & 1.904 & 1.907 & 0.111 & 0.107 \\ \hline
    \end{tabular}%
  \label{tab:m-diag}%
\end{table}%

Although CAS-PDFT is a method designed for multireference calculations, it is important to test whether tPBE can determine accurate geometries for systems with predominantly single-reference character.  Ten different molecules were considered, and the resultant geometric parameters are available in Table~S1.  The M-diagnostic~\cite{Mdiag} was used to evaluate the degree of multireference character; these results are given in Table~\ref{tab:m-diag}.  Generally, systems exhibiting $M$ values greater than 0.1 are considered to have significant multireference character.  Acrolein, butadiene, pyridine, and maleic anhydride all were found to have M-diagnostic values slightly larger than 0.1.

In order to evaluate the performance of each method, each converged structure was best-fit (via rigid rotation and translation) to the corresponding reference structure from the SE47 database by minimizing the root mean square~(RMS) distance between the atomic centers.  Results using the cc-pVDZ basis set are given in Figure~\ref{fig:ccd}, and results using the cc-pVTZ basis set are given in Figure~\ref{fig:cct}.  The largest distance discrepancies between PBE and tPBE are seen for the four molecules exhibiting significant multireference character.  In those cases, tPBE shows a noticeable improvement over PBE.  Increasing the basis set size lowers the RMS displacement for tPBE, CASPT2, and PBE while the CASSCF RMS displacements become larger for several molecules. 

The selection of active spaces can possibly affect the quality of results for certain bonds and angles.  For example, C-H bonding and antibonding orbitals are often not included in the active space of the largest systems due to active space size limitations.  This is reflected in the resultant tPBE bond lengths--the C--H and N--H bonds tend to have larger errors than for the C--C, N--C, and C--O bonds.  In the case of pyridine, a (6,6) full-$\pi$ active space was selected.  The errors in the tPBE bond lengths (cc-pVTZ) are about 0.001~\r{A} for the C--C and C--N bonds, but the C--H bond errors are near 0.010~\r{A}.  A similar trend can be seen in the CASSCF results.  CASPT2 appears less susceptible to active space limitations as its errors are quite similar across all bonds.  Because CAS-PDFT does not improve or alter the reference wave function~(it only supplies an energy correction), the reference wave function likely has the potential to exhibit greater influence over the quality of the CAS-PDFT result than in a method such as CASPT2, which computes a first-order correction to the reference wave function.

A total of 74 geometric parameters (39~bond lengths and 35~bond angles) were optimized.  If we compare the tPBE, CASSCF, and CASPT2 results (which all employ the same active space), we see that the tPBE results are more similar to the CASPT2 results than to the CASSCF results for 59 of these data when using the cc-pVDZ basis set and 57 of these data when using the cc-pVTZ basis set.  Additionally, if we view the tPBE and CASPT2 methods as ``correcting'' the CASSCF result,  tPBE and CASPT2 agree on the direction of the change in the geometric parameter for 63 of the geometrical variables when using the cc-pVDZ and 64 of them when using the cc-pVTZ basis set.  A comparison between the tPBE and PBE results shows that both methods predict very similar geometries for most species.  In the majority of bond distances and bond angles, tPBE gives a more accurate result than PBE, with the most pronounced differences seen in acrolein, butadiene, pyridine, and maleic anhydride.  

\begin{table}[htbp]
  \caption{Mean signed error (MSE) and mean unsigned error (MUE) over the set of 39 bond lengths and 35 bond angles.  Errors are calculated relative to the SE47 database~\cite{Barone}.} 
    \begin{tabular}{lcccccccc}\hline \hline
          & \multicolumn{4}{c}{cc-pVDZ} & \multicolumn{4}{c}{cc-pVTZ} \\ \cmidrule(lr){2-5} \cmidrule(lr){6-9}
          & tPBE & CASSCF & CASPT2 & PBE & tPBE & CASSCF & CASPT2 & PBE \\ \cmidrule(lr){2-5} \cmidrule(lr){6-9}
    \textbf{MSE} & & & & & & & & \\     \cline{1-1}
    bonds lengths & 0.014 & 0.005 & 0.013 & 0.016 & 0.006 & -0.003 & 0.001 & 0.008 \\
    bond angles & -0.1  & -0.1  & -0.1  & -0.3  & 0.0   & 0.0   & -0.3  & 0.0 \\
    &&&&&&&& \\
    \textbf{MUE}      &       &       &       &       &       &       &       &  \\ \cline{1-1}

    bonds lengths & 0.014 & 0.008 & 0.013 & 0.016 & 0.007 & 0.008 & 0.003 & 0.009 \\
    bond angles & 0.4   & 0.6   & 0.6   & 0.7   & 0.3   & 0.6   & 0.5   & 0.4 \\ \hline
    \end{tabular}%
  \label{tab:errors}%
\end{table}%

The mean signed error~(MSE) and mean unsigned error~(MUE) over the set of 39 bond lengths and 35 bond angles for each method are given in Table~\ref{tab:errors}.  The errors for all methods are small.  The MC-PDFT results obtained with tPBE show similar MSEs and MUEs for both basis sets.  Notably, the tPBE results show an improvement upon the KS-DFT results obtained with PBE.  Further, the tPBE results parallel the CASPT2 and PBE results in terms of performance in going from the smaller cc-pVDZ to the larger cc-pVTZ basis set, obtaining greater accuracy with the larger basis set.  In contrast, the MUE for the CASSCF method does not show an improvement upon going to a larger basis set.

\section{Conclusions}
In this work we have derived the working equations necessary to calculate analytical gradients for the CAS-PDFT method.  We have used these routines to perform geometry optimizations on a set of ten molecules, including seven single-reference systems and three strongly correlated ones.  We have found that CAS-PDFT with the tPBE functional performs comparably to CASPT2 on these predominantly-single-referenced systems.  In future work we apply the analytical gradient treatment of CAS-PDFT to more multireferenced systems, including transition states and excited states.

Further improvement of CAS-PDFT geometries is also possible through the implementation of new on-top pair density functionals.  Our tPBE results parallel the KS-DFT results with PBE, and PBE is often not the best functional for geometry optimizations~\cite{yu_2016}.

The current implementation for analytical CAS-PDFT gradients is restricted to state-specific CASSCF reference wave functions.  In future work we plan to develop state-averaged CAS-PDFT analytic gradient routines.

\section{Associated Content}
\subsection{Supporting Information}
The supporting information is available free of charge on the ACS Publications website at DOI: 10.1021/

The Supporting information contains optimized Cartesian coordinates and some examples showing results for ftPBE.

\section{Acknowledgment}
This work was supported in part by the National Science Foundation by grants CHE-1464536.  K.M.K. acknowledges support from summer research fellowship CHE-1359181.  R.L. acknowledges support by the Swedish Research Council (Grant 2016-03398).

\appendix
\section{Appendix: Analytic gradients for fully-translated functionals}
In the fully-translated (``ft'') functional scheme, the on-top functional depends on the derivative of the on-top pair density $\Pi'$.  The functional transformation becomes $E_{\mathrm{ot}}[\rho,\Pi,\rho',\Pi'] = E_{\mathrm{xc}}[\tilde{\rho}_{\alpha},\tilde{\rho}_{\beta},\tilde{\rho}_{\alpha}',\tilde{\rho}_{\beta}']$, where the ft-densities are written as~\cite{mcpdft_ft} 
\begin{align}
 \tilde{\rho}_{\alpha}({\bf{r}}) &= 
 \begin{cases}
 \frac{\rho({\bf{r}})}{2} (1+\zeta_t(\bf{r})) &  R({\bf{r}}) < R_0 \\
 \frac{\rho({\bf{r}})}{2} (1+\zeta_{ft}({\bf{r}})) & R_0 \leq R({\bf{r}}) \leq R_1  \\
 \frac{\rho({\bf{r}})}{2}  & R({\bf{r}}) > R_1
 \end{cases} \label{eq:ftrans1} \\
 \tilde{\rho}_{\beta}({\bf{r}}) &= 
  \begin{cases}
 \frac{\rho({\bf{r}})}{2} (1-\zeta_t(\bf{r})) &  R({\bf{r}}) \leq R_0 \\
 \frac{\rho({\bf{r}})}{2} (1-\zeta_{ft}({\bf{r}})) & R_0 \leq R({\bf{r}}) \leq R_1  \\
 \frac{\rho({\bf{r}})}{2}  & R({\bf{r}}) > R_1
 \end{cases} \\
 \tilde{\rho}_{\alpha}'({\bf{r}}) &= 
  \begin{cases}
 \frac{\rho'({\bf{r}})}{2} (1+\zeta_t({\bf{r}})) + \frac{\rho({\bf{r}})}{2}\zeta_{t}'({\bf{r}}) &  R({\bf{r}}) \leq R_0 \\
 \frac{\rho'({\bf{r}})}{2} (1+\zeta_{ft}({\bf{r}})) + \frac{\rho({\bf{r}})}{2}\zeta_{ft}'({\bf{r}}) & R_0 \leq R({\bf{r}}) \leq R_1  \\
 \frac{\rho'({\bf{r}})}{2}  & R({\bf{r}}) > R_1
 \end{cases} \\
 \tilde{\rho}_{\beta}'({\bf{r}}) &= 
  \begin{cases}
 \frac{\rho'({\bf{r}})}{2} (1-\zeta_t({\bf{r}})) - \frac{\rho({\bf{r}})}{2}\zeta_{t}'({\bf{r}}) &  R({\bf{r}}) \leq R_0 \\
  \frac{\rho'({\bf{r}})}{2} (1-\zeta_{ft}({\bf{r}})) - \frac{\rho({\bf{r}})}{2}\zeta_{ft}'({\bf{r}}) & R_0 \leq R({\bf{r}}) \leq R_1  \\
 \frac{\rho'({\bf{r}})}{2}  & R({\bf{r}}) > R_1
 \end{cases} \label{eq:ftrans4}
\end{align}
where $\zeta_{t}(\bf{r})$, $\zeta_{ft}(\bf{r})$,$\zeta_{t}'(\bf{r})$ and $\zeta_{ft}'(\bf{r})$ are defined as
\begin{align}
\zeta_t(\bf{r}) &= \sqrt{1-R(\bf{r})} \\
\zeta_{ft}(\bf{r}) &= A(R({\bf{r}}) - R_1)^5 + B(R({\bf{r}}) - R_1)^4 + C(R({\bf{r}}) - R_1)^3 \\
\zeta_t'(\bf{r}) &= -\frac{1}{2} \frac{R'(\bf{r})}{\zeta_t(\bf{r})} \\
\zeta_{ft}'(\bf{r}) &= R'({\bf{r}}) [5A(R({\bf{r}})-R_1)^4 +4B(R({\bf{r}})-R_1)^3 + 3C(R({\bf{r}})-R_1)^2]
\end{align}
with the following parameters:
\begin{align}
R_0 &= 0.9 \\
R_1 &= 1.15 \\
A &= -475.60656009 \\
B &= -379.47331922 \\
C &= -85.38149682
\end{align}
The gradient $R'({\bf{r}})$ is written as
\begin{equation}
R'({\bf{r}}) = \frac{\Pi'({\bf{r}})}{[\rho({\bf{r}})/2]^2} - \frac{\Pi({\bf{r}}) \rho'({\bf{r}}) }{[\rho({\bf{r}})/2]^3}
\end{equation}
The evaluation of the one-electron and two-electron on-top potentials $V_{pq}$ (Eq.~\ref{eq:oeot}) and $v_{pqrs}$ (Eq.~\ref{eq:teot}) require the derivatives of Eqs.~\ref{eq:ftrans1}-\ref{eq:ftrans4} with respect to the one- and two-body density matrices.  Derivatives with respect to the one-body density matrix are given by
\begin{align}
 \frac{\partial \tilde{\rho}_{\alpha}({\bf r})}{\partial D_{pq}} &= 
 \begin{cases} 
 \frac{(1+\zeta_t(\bf{r}))}{2}\frac{\partial \rho({\bf{r}})}{\partial D_{pq}} + \frac{\rho({\bf{r}})}{2} \frac{\partial\zeta_t(\bf{r}))}{\partial D_{pq}} &  R({\bf{r}}) < R_0 \\
\frac{(1+\zeta_{ft}(\bf{r}))}{2}\frac{\partial \rho({\bf{r}})}{\partial D_{pq}} + \frac{\rho({\bf{r}})}{2} \frac{\partial\zeta_{ft}(\bf{r}))}{\partial D_{pq}} & R_0 \leq R({\bf{r}}) \leq R_1  \\
  \frac{1}{2} \frac{\partial\rho({\bf{r}})}{\partial D_{pq}} & R({\bf{r}}) > R_1
 \end{cases} \\
  \frac{\partial \tilde{\rho}_{\beta}({\bf r})}{\partial D_{pq}} &= 
  \begin{cases}
 \frac{(1-\zeta_t(\bf{r}))}{2}\frac{\partial \rho({\bf{r}})}{\partial D_{pq}} - \frac{\rho({\bf{r}})}{2} \frac{\partial\zeta_t(\bf{r}))}{\partial D_{pq}} &  R({\bf{r}}) < R_0 \\
\frac{(1-\zeta_{ft}(\bf{r}))}{2}\frac{\partial \rho({\bf{r}})}{\partial D_{pq}} - \frac{\rho({\bf{r}})}{2} \frac{\partial\zeta_{ft}(\bf{r}))}{\partial D_{pq}} & R_0 \leq R({\bf{r}}) \leq R_1  \\
  \frac{1}{2} \frac{\partial\rho({\bf{r}})}{\partial D_{pq}} & R({\bf{r}}) > R_1
 \end{cases} \\
 \frac{\partial \tilde{\rho}_{\alpha}'({\bf r})}{\partial D_{pq}} &= 
 \begin{cases}
   \frac{(1+\zeta_t(\bf{r}))}{2} \frac{\partial \rho'({\bf{r}})}{\partial D_{pq}} + \frac{\rho'({\bf{r}})}{2} \frac{\partial\zeta_t(\bf{r})}{\partial D_{pq}} + \frac{\rho({\bf{r}})}{2} \frac{\partial\zeta_t'(\bf{r})}{\partial D_{pq}} + \frac{\zeta_t'(\bf{r})}{2} \frac{\partial \rho({\bf{r}})}{\partial D_{pq}}  &  R({\bf{r}}) < R_0 \\
   \frac{(1+\zeta_{ft}(\bf{r}))}{2} \frac{\partial \rho'({\bf{r}})}{\partial D_{pq}} + \frac{\rho'({\bf{r}})}{2} \frac{\partial\zeta_{ft}(\bf{r})}{\partial D_{pq}} + \frac{\rho({\bf{r}})}{2} \frac{\partial\zeta_{ft}'(\bf{r})}{\partial D_{pq}} + \frac{\zeta_{ft}'(\bf{r})}{2} \frac{\partial \rho({\bf{r}})}{\partial D_{pq}} & R_0 \leq R({\bf{r}}) \leq R_1  \\
  \frac{1}{2} \frac{\partial\rho'({\bf{r}})}{\partial D_{pq}} & R({\bf{r}}) > R_1
 \end{cases}  \\
 \frac{\partial \tilde{\rho}_{\beta}'({\bf r})}{\partial D_{pq}} &= 
 \begin{cases}
   \frac{(1-\zeta_t(\bf{r}))}{2} \frac{\partial \rho'({\bf{r}})}{\partial D_{pq}} - \frac{\rho'({\bf{r}})}{2} \frac{\partial\zeta_t(\bf{r})}{\partial D_{pq}} - \frac{\rho({\bf{r}})}{2} \frac{\partial\zeta_t'(\bf{r})}{\partial D_{pq}} - \frac{\zeta_t'(\bf{r})}{2} \frac{\partial \rho({\bf{r}})}{\partial D_{pq}}  &  R({\bf{r}}) < R_0 \\
   \frac{(1-\zeta_{ft}(\bf{r}))}{2} \frac{\partial \rho'({\bf{r}})}{\partial D_{pq}} - \frac{\rho'({\bf{r}})}{2} \frac{\partial\zeta_{ft}(\bf{r})}{\partial D_{pq}} - \frac{\rho({\bf{r}})}{2} \frac{\partial\zeta_{ft}'(\bf{r})}{\partial D_{pq}} - \frac{\zeta_{ft}'(\bf{r})}{2} \frac{\partial \rho({\bf{r}})}{\partial D_{pq}} & R_0 \leq R({\bf{r}}) \leq R_1  \\
  \frac{1}{2} \frac{\partial\rho'({\bf{r}})}{\partial D_{pq}} & R({\bf{r}}) > R_1
 \end{cases}  
 \end{align}
where we employ the following intermediate derivatives:
\begin{align}
\frac{\partial \rho({\bf{r}})}{\partial D_{pq}} &= \phi_p({\bf{r}})\phi_q({\bf{r}}) \\ 
\frac{\partial \rho'({\bf{r}})}{\partial D_{pq}} &= \phi'_p({\bf{r}})\phi_q({\bf{r}}) +  \phi_p({\bf{r}})\phi'_q({\bf{r}}) \\
\frac{\partial\zeta_t(\bf{r})}{\partial D_{pq}} &= -\frac{1}{2\zeta_t({\bf{r}})} \frac{\partial R({\bf{r}})}{\partial D_{pq}} \\
\frac{\partial R({\bf{r}})}{\partial D_{pq}} &= -\frac{8 \Pi({\bf{r}})}{[\rho({\bf{r}})]^3} \phi_p({\bf{r}}) \phi_q({\bf{r}}) \\
\frac{\partial\zeta'_t(\bf{r})}{\partial D_{pq}} &= \frac{R'({\bf{r}})}{2 [\zeta_t({\bf{r}})]^2} \frac{\partial\zeta_t(\bf{r})}{\partial D_{pq}} - \frac{1}{2 \zeta_t({\bf{r}})} \frac{\partial R'({\bf{r}})}{\partial D_{pq}} \\
\frac{\partial R'({\bf{r}})}{\partial D_{pq}} &= \bigg{(} \frac{4\Pi'({\bf{r}})}{[\rho({\bf{r}})]^3}  + \frac{24 \Pi({\bf{r}})}{[\rho({\bf{r}})]^4} \bigg{)} \frac{\partial \rho({\bf{r}})}{\partial D_{pq}} - \frac{8 \Pi ({\bf{r}})}{[\rho({\bf{r}})]^3} \frac{\partial \rho'({\bf{r}})}{\partial D_{pq}} \\
\frac{\partial\zeta_{ft}(\bf{r})}{\partial D_{pq}} &= [5A(R({\bf{r}})-R_1)^4 +4B(R({\bf{r}})-R_1)^3 + 3C(R({\bf{r}})-R_1)^2] \frac{\partial R({\bf{r}})}{\partial D_{pq}}  \\
\frac{\partial\zeta'_{ft}(\bf{r})}{\partial D_{pq}} &= \big{[}20A(R({\bf{r}})-R_1)^3 +12B(R({\bf{r}})-R_1)^2 + 6C(R({\bf{r}})-R_1)\big{]} R'({\bf{r}}) \frac{\partial R({\bf{r}})}{\partial D_{pq}} \nonumber \\
 &+ [5A(R({\bf{r}})-R_1)^4 +4B(R({\bf{r}})-R_1)^3 + 3C(R({\bf{r}})-R_1)^2] \frac{\partial R'({\bf{r}})}{\partial D_{pq}}
\end{align}
The derivatives with respect to the two-body density matrix are given by
\begin{align}
 \frac{\partial \tilde{\rho}_{\alpha}({\bf r})}{\partial d_{pqst}} &= 
 \begin{cases} 
\frac{\rho({\bf{r}})}{2} \frac{\partial \zeta_t({\bf{r}})}{\partial d_{pqst}}&  R({\bf{r}}) < R_0 \\
\frac{\rho({\bf{r}})}{2} \frac{\partial \zeta_{ft}({\bf{r}})}{\partial d_{pqst}}& R_0 \leq R({\bf{r}}) \leq R_1  \\
0 & R({\bf{r}}) > R_1
 \end{cases} \\
  \frac{\partial \tilde{\rho}_{\beta}({\bf r})}{\partial d_{pqst}} &= 
  \begin{cases}
-\frac{\rho({\bf{r}})}{2} \frac{\partial \zeta_t({\bf{r}})}{\partial d_{pqst}}&  R({\bf{r}}) < R_0 \\
-\frac{\rho({\bf{r}})}{2} \frac{\partial \zeta_{ft}({\bf{r}})}{\partial d_{pqst}}& R_0 \leq R({\bf{r}}) \leq R_1  \\
0 & R({\bf{r}}) > R_1
 \end{cases} \\
 \frac{\partial \tilde{\rho}_{\alpha}'({\bf r})}{\partial d_{pqst}} &= 
 \begin{cases}
\frac{\rho({\bf{r}})}{2} \frac{\partial \zeta'_t({\bf{r}})}{\partial d_{pqst}} + \frac{\rho'({\bf{r}})}{2} \frac{\partial \zeta_t({\bf{r}})}{\partial d_{pqst}}  &  R({\bf{r}}) < R_0 \\
\frac{\rho({\bf{r}})}{2} \frac{\partial \zeta'_{ft}({\bf{r}})}{\partial d_{pqst}} + \frac{\rho'({\bf{r}})}{2} \frac{\partial \zeta_{ft}({\bf{r}})}{\partial d_{pqst}}  & R_0 \leq R({\bf{r}}) \leq R_1  \\
0 & R({\bf{r}}) > R_1
 \end{cases}  \\
 \frac{\partial \tilde{\rho}_{\beta}'({\bf r})}{\partial d_{pqst}} &= 
 \begin{cases}
-\frac{\rho({\bf{r}})}{2} \frac{\partial \zeta'_t({\bf{r}})}{\partial d_{pqst}} - \frac{\rho'({\bf{r}})}{2} \frac{\partial \zeta_t({\bf{r}})}{\partial d_{pqst}}  &  R({\bf{r}}) < R_0 \\
-\frac{\rho({\bf{r}})}{2} \frac{\partial \zeta'_{ft}({\bf{r}})}{\partial d_{pqst}} - \frac{\rho'({\bf{r}})}{2} \frac{\partial \zeta_{ft}({\bf{r}})}{\partial d_{pqst}}  & R_0 \leq R({\bf{r}}) \leq R_1  \\
0 & R({\bf{r}}) > R_1
 \end{cases}  
 \end{align}
 where we have used the intermediates
 \begin{align}
 \frac{\partial \zeta_{t}({\bf{r}})}{\partial d_{pqst}} &= -\frac{1}{2  \zeta_{t}({\bf{r}})} \frac{\partial R({\bf{r}})}{\partial d_{pqst}} \\
 \frac{\partial R({\bf{r}})}{\partial d_{pqst}} &= \frac{4}{[\rho({\bf{r}})]^2} \frac{\partial \Pi({\bf{r}})}{\partial d_{pqst}} \\
\frac{\partial \Pi({\bf{r}})}{\partial d_{pqst}} &= \phi_p({\bf{r}})\phi_q({\bf{r}})\phi_s({\bf{r}})\phi_t({\bf{r}}) \\
 \frac{\partial \zeta'_{t}({\bf{r}})}{\partial d_{pqst}} &= -\frac{1}{2} \bigg{(} \frac{-R'({\bf{r}})}{[\zeta_t({\bf{r}})]^2} \frac{\partial \zeta_{t}({\bf{r}})}{\partial d_{pqst}} + \frac{1}{\zeta_{t}({\bf{r}})} \frac{\partial R'({\bf{r}})}{\partial d_{pqst}} \bigg{)}   \\
  \frac{\partial R'({\bf{r}})}{\partial d_{pqst}} &= \frac{4}{[\rho({\bf{r}})]^2} \frac{\partial \Pi'({\bf{r}})}{\partial d_{pqst}} - \frac{8 \rho'({\bf{r}})}{[\rho({\bf{r}})]^3} \frac{\partial \Pi({\bf{r}})}{\partial d_{pqst}}  \\
 \frac{\partial \Pi'({\bf{r}})}{\partial d_{pqst}} &= \phi'_p({\bf{r}})\phi_q({\bf{r}})\phi_s({\bf{r}})\phi_t({\bf{r}}) + \phi_p({\bf{r}})\phi'_q({\bf{r}})\phi_s({\bf{r}})\phi_t({\bf{r}})\nonumber \\ 
 &+ \phi_p({\bf{r}})\phi_q({\bf{r}})\phi'_s({\bf{r}})\phi_t({\bf{r}}) + \phi_p({\bf{r}})\phi_q({\bf{r}})\phi_s({\bf{r}})\phi'_t({\bf{r}}) \\
 \frac{\partial \zeta_{ft}({\bf{r}})}{\partial d_{pqst}} &= [5A(R({\bf{r}})-R_1)^4 +4B(R({\bf{r}})-R_1)^3 + 3C(R({\bf{r}})-R_1)^2] \frac{\partial R({\bf{r}})}{\partial d_{pqst}}  \\
 \frac{\partial \zeta'_{ft}({\bf{r}})}{\partial d_{pqst}} &=  \big{[}20A(R({\bf{r}})-R_1)^3 +12B(R({\bf{r}})-R_1)^2 + 6C(R({\bf{r}})-R_1)\big{]} R'({\bf{r}})  \frac{\partial R({\bf{r}})}{\partial d_{pqst}}\nonumber \\
 &+ [5A(R({\bf{r}})-R_1)^4 +4B(R({\bf{r}})-R_1)^3 + 3C(R({\bf{r}})-R_1)^2] \frac{\partial R'({\bf{r}})}{\partial d_{pqst}}\\
 \end{align}

\bibliography{ref}
\begin{tocentry}
\begin{center}
\includegraphics[scale=1]{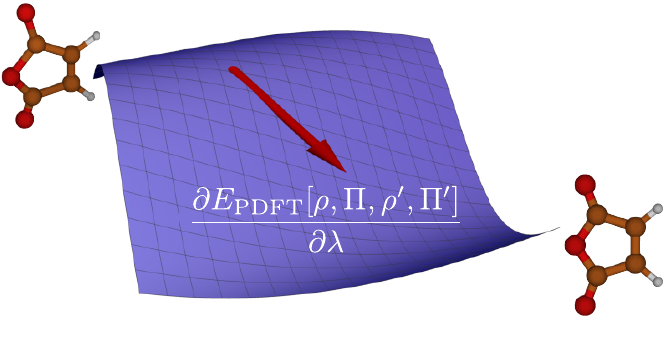}
\end{center}
\end{tocentry}

\end{document}